\documentclass{article}
\def\sa{{\vspace{1mm}{\em Sal -- }}}
\def\si{{\vspace{1mm}{\em Simp -- }}}

\begin{document}

\section*{A dialog on quantum gravity\\[2mm]
\normalsize Carlo Rovelli}

\parindent=0pt
\parskip=.1cm

{\em The debate between loop quantum gravity and string theory is
sometime lively, and it is hard to present an impartial view on the
issue.  Leaving any attempt to impartiality aside, I report here,
instead, a conversation on this issue, overheard in the cafeteria of a
Major American University.  The personae of the dialog are Professor
Simp, a high energy physicist, and a graduate student, Sal.  The
Professor has heard that Sal has decided to work in loop gravity, and
gently tries to talk her out.  Here is what was heard.} 
\vskip.8cm

\sa Hi Prof. 

\si Hi Sal. So, I hear you are interested in loops. 

\sa Yes indeed, have been reading. 

\si And?

\sa Like it. 

\si Do you want to start doing loops? 

\sa Maybe. 

\si You are not going to find a job ... 

\sa Maybe.  I prefer follow what fascinates me 
in science.  Time to grow old later.

\si Hmm.  And what fascinates you so much? 

\sa The merging of GR and QM. Understanding space and time, that
stuff.

\si String theory merges GR and QM. 

\sa Yes, but the price is too high. 

\si Too high?

\sa Extra dimensions, supersymmetry, infinite fields \ldots

\si This is not a price, is fascinating new physics. 

\sa For the moment, is not new physics, is just our 
speculations. 

\si  So is loop quantum gravity. 

\sa Yes of course, but loop gravity does the trick using only GR and
QM, and this we know for good, without all the extra baggage.

\si Really does the trick? All works with loops? 

\sa No, much is missing.  But much is missing in string 
theory as well. 

\si Not so much.  In string theory you can compute
scattering amplitudes and cross sections.  I think you can't even do
that in loop gravity.

\sa True, but there are other things you can compute, which
you can't compute in string theory.

\si That is?  

\sa Spectra of areas and volume, for instance.

\si But you cannot measure that. 

\sa You can in principle\ldots

\si Maybe in principle, but not in practice \ldots 

\sa Nor you can measure, in practice, the cross sections 
predicted by string theory \ldots \  But there is a big difference.

\si That is?

\sa The prediction with loops are unique, well fixed.  Maybe
up to a parameter, but no more than that.  If, or when, we are able to
measure areas, say cross sections, with Planck scale precision, we
will find the numbers predicted by loop gravity.  Or not.  And we will
know the theory is right.  Or wrong.  This is good science.  Isn' it?

\si There are predictions from string theory as well. 

\sa Like?

\si Like large extra dimensions. Supersymmetry. 
Transitions that cannot happen in the standard model. 

\sa You mean that if we do not find the experimental
consequences of large extra dimensions we conclude string theory is
wrong?

\si Of course not, large extra dimensions exist only in very special 
models.  

\sa So, the experiments on extra dimensions cannot kill
string theory?

\si No they cannot. 

\sa And if supersymmetry is not found at the scale we expect,
we do not abandon string theory?

\si No we don't.  It will be at a higher scale.

\sa So, which experiment could kill string theory, in 
principle? 

\si Nothing I could think off. The theory is very strong. 

\sa Seems to me is very weak.  A good scientific theory is a
theory that can be falsified. 

\si I am not a philosopher \ldots 

\sa I mean, it is a theory that gives definite predictions. 
Not a theory that never tells us what we will see in the next
experiment, and that can accommodate any outcome of any experiment. 
What's the good of a theory that can accommodate anything and the
contrary of anything?

\si You are exaggerating a bit \ldots 

\sa Yes\ldots but at least, is there a version of string theory
that agrees with the reality as we see it?

\si Of course!  There are Yang-Mills fields, quarks, the
graviton!  What do you mean?

\sa I mean a version of the theory and its vacuum, a
Calabi-Yau manifold, or some other way of breaking down the theory to
4d and getting precisely the standard model, with the masses and the
particle content we see in the world, including the families?

\si I think that there are Calabi-Yau manifolds that give a physics 
quite similar to the standard model. 

\sa Okay, but is there one that give precisely the standard model 
physics in the regime we have tested it? 

\si Hmm\ldots No, I do not think so\ldots At least not precisely
\ldots

\sa So, so far string theory does not really agree with the world we
see \ldots\ requires a long list of very complicated things that we do
not see, such as supersymmetry and extradimensions \ldots\ and does
not give any definite, univocal, prediction on future experiments
\ldots\ is this a serious theory \ldots ?

\si It is only because we are not able to truly compute with
it.

\sa Of course, but this is cheap, it can be said about any
sufficiently complicated theory \ldots\ why should I believe strings
in particular?

\si Because it is the only theory we have for combining GR and QM.
Because it gives a finite quantum theory, including gravity.  Because
it deals with what is missing in the standard model.  Because many
aspects of reality that have been confirmed from experiment follow
from string theory: gravity, gauge theories, fermions\ldots.  Because
the theory has only one free parameter instead of the 19 of the
standard model.  Because it brings everything together, it is the
theory of everything.  And because it is a very beautiful theory.

\sa Prof, can we discuss all that during lunch? 

\si Sure, if you then accept discussing loop gravity. 

\vskip.5cm

{\em During this exchange Sal and Professor Simp were standing in line and
getting food.  At this point, they sit at a table.  Sal has noted down
a few words on a napkin.  A few other students, curious of the
exchange, sit nearby and listen. }

\vskip.5cm

\sa I have noted your points for string theory.  Let me start by
saying that obviously string theory is an extremely remarkable theory,
it is full of wonders and surprises, it touches so much mathematics,
it clearly talks well with the physics we know.  I have a tremendous
respect for the people who have developed it, who are my heroes.  But
the issue I want to address is to which extent we can be sure, at
present, that this is the correct theory for describing nature.  If we
are reasonably sure of that, there is no reason to study alternatives. 
If we are not, we better study alternatives.  Especially given that
alternatives exist.  So, let me address your list of reasons to
believe that string theory is physically correct.  I start with the
first: it is the only theory we have for combining GR and QM\ldots

\si I take this point back for the moment.  I forgot you are studying
loops.  There is of course loop theory as well.  Let's talk loops
later.

\sa Good.  You say that the theory has only one parameter, instead of
the 19 of the standard model.  In general, we are interested in a
theory with less parameters if this theory allows us to derive, to
compute the old free parameters.  Can we do that using string theory?

\si Not yet. 

\sa So, as far as the 19 parameters of the standard model are
concerned, string theory has basically replaced an open problem with
another open problem.

\si Yes, but there is a hope to solve it. 

\sa I have no doubts that if all the wishful thinking of the string
theorists is realized, then string theory is a perfect theory, and we
do not need loops.  But also if all the wishful thinking of the loop
people is realized, then loop theory is a perfect theory, and we do
not need strings.  Let's not talk about hopes, let's talk about
achievements.

\si Then the 19 parameters of the standard model are not understood.

\sa All right.  There are other open problems of the standard model,
besides being able to compute all the constants.  Like understanding
why there are three families.  Does string theory solve that?

\si \ldots no \ldots

\sa why the cosmological constant is small ? 

\si \ldots no \ldots

\sa giving a better account of symmetry breaking ?

\si \ldots no \ldots

\sa so? 

\si Well, the theory derives the full complexity of the standard model
from an extremely simple picture \ldots

\sa Wait a minute.  I agree the bosonic string is a simple physical
picture.  But the bosonic string is physically bad because of the
tachyon and it definitely does not give the standard model.  You have
to go, say, to the heterotic string, with gauge groups,
superfields\ldots\ a different behavior of the two halves of the
theory \ldots\ I wouldn't call this an extremely simple picture. 
Plus, you have to select a very specific and complicated internal
space, by hand, so far, to get the standard model.  In general, you
have to get to the second volume of a book on string theory to just
begin to be able to understand the definition of models that have a
chance of being realistic \ldots\ and with all that, we haven't even
found a way to derive the standard model with all its details\ldots

\si You talk as if string theory was an extremely complicated
machinery invented out of the blue for no reason.  Its funny features,
like extradimensions, supersymmetries and all that, were not put there
just for fun.  There has been a compelling logical evolution.  It is
remarkable that these complications have solved specific theoretical
problems combining into the final theory\ldots

\sa Which problems? \ldots

\si \ldots it all started with the dual models and the Veneziano
amplitude\ldots

\sa \ldots the amplitude Gabriele Veneziano wrote down realizes the
duality that was conjectured to hold in the strong interaction
physical amplitudes between the $s$ channel and the $t$ channel\ldots

\si You know your history.

\sa And does the Veneziano formula describe correctly the cross
sections that we observe?

\si No, it does not.  The physical high energy behavior of the strong
interactions cross sections is not the one predicted by this formula. 

\sa Therefore a good physicist concludes that Veneziano formula was a 
nice theoretical idea, but not one that Nature likes, and abandons it, 
to study something else \ldots\ I suppose we have to believe Nature, 
not to believe our beautiful formulas when they disagree \ldots\ 

\si In fact, the Veneziano formula was abandoned for strong
interactions, but so much was found out of it, it was understood it
could emerge from a string theory and it was realized that it could be
used in a much better way.  

\sa Wait a minute, this story gives a historical account of the birth
of an idea.  But it has nothing to do with providing reasons for
believing that this idea is physically correct.  If anything, it shows
that it was a wrong idea to start with.  The Veneziano amplitude and
the Dolen-Horn-Schmid duality in strong interactions were motivated by
the observation of the resonances with high spin, the rough
proportionality between mass and spin\ldots\ If I understand
correctly, all this is today understood on the basis of QCD. Even the
apparent ``stringy" behavior can be understood from QCD; tubes of flux
of the color lines of force behave like small strings in some
approximation.  Therefore it is reasonable that some sort of string
theory gives an approximate account of the phenomenology.  The correct
physical conclusion is that string theory is an approximate
description at some scale, not a fundamental theory.

\si This in fact was the conclusion.  However, it turned out that the
Veneziano formula opens a vast and beautiful theoretical world.  So
much came out of it.  Strings are good description for other gauge
theories, like N=4 Super-Yang-Mills, and it is probable that a string
description will be found for large N QCD.

\sa I am reading a book, written long ago, where I just found this
phrase: \emph{``\ldots perch\`e i nostri discorsi hanno a essere sopra
un mondo sensibile, e non sopra un mondo di carta."} Roughly:
\emph{``\ldots our arguments have to be about the world we experience,
not about a world made of paper".} None of those theories is connected
to the world, as far as we know.

\si The theory naturally includes a graviton, and we do experience
gravity, and the theory is finite at high energy.

\sa But it is only consistent in 26 dimensions, unlikely the world we 
experience. 

\si But this can be corrected by using Kaluza-Klein compactification.

\sa And we get a theory with a tachyon, that we do not experience, and
questions the consistency of the theory.

\si This can be corrected with supersymmetry. 

\sa and so on\ldots\ You keep getting a theory that is either
inconsistent or badly disagrees with reality, and you keep making it
more and more complicated\ldots 

\si \ldots until you get a theory that has a chance of being
consistent and agreeing with nature \ldots

\sa Maybe, but the different ingredients of the theory are not
solutions of problems of the standard model, or solutions of problems
we have in understanding of the world: they are solutions to problems
raised by other ingredients of the same theory.  According to the
Catholic doctrine there are two miracles happening in a Mass: the
first miracle is that wine becomes truly blood.  The second miracle
is that the blood looks and smells like wine\ldots \ it is just a
miracle added to patch up the inconsistency created by the first
\ldots

\si Don't be irreverent. The point is that, you continue until you 
finally get a theory that is consistent and agrees with nature. 

\sa \ldots or until you get a theory where you cannot compute
anything, you cannot reduce it down to 4d, you cannot get the Standard
Model back \ldots\ and it is sufficiently involved and unfathomable
that you can't anymore prove it wrong, and just declare that
everything is in it, to be found by future generations\ldots\ And if
any problem arise --say the vacua that you have found so far are all
unstable, as was found in a recent paper by Gary Horowitz and others
-- the theory is sufficiently complicated that you can always
wishfully think that there is something else in it that can save the
day \ldots

\si It is not the fault of the theoretical physicist if the path of
the natural evolution of the research has lead to a theory which is
very complicated.

\sa And if it \emph{was} the fault of the theoretical physicist?  I
suppose when you say ``the path of the natural evolution of the
research" you mean the line that goes along Fermi theory, QED,
$SU(2)\times U(1)$, QCD, the standard model, and then grand unified
theories, the revival of Kaluza-Klein, supersymmetry, supergravity,
\ldots\ strings\ldots

\si Yes. 

\sa But what if this ``path of natural evolution" has taken a wrong
turn at some point.  Seems to me there is precise break along this
path.

\si What do you mean? 

\sa Dirac predicted the positron, and it was found.  Feynman and
friends developed a calculation method for photon-electron
interactions, and it works to devastating precision.  Weinberg Glashow
and Salam predicted the neutral currents and they were found, and the
W and Z particles and Carlo Rubbia found them, precisely where
predicted, just to name some \ldots

\si So? 

\sa And then? 

\si Then what? 

\sa Then the Veneziano formula predicted a very soft high energy
behavior of the amplitudes, and nature was \emph{not} like that.  The
grand unified theories predicted proton decay at some precise scale,
and proton decay was \emph{not} found where expected.  Kaluza-Klein
theory, revived, predicted the existence of a scalar field that was
searched by Dicke, and \emph{not} found.  Supersymmetry predicted the
supersymmetric particles and these were \emph{not} found where
repeatedly annunciated.  Extra dimensions did \emph{not} show up where
recently suggested by string theory\ldots

\si But the proton may take a bit longer to decay, the
masses of the supersymmetric partners may be higher \ldots

\sa Of course, they ``might".  Everything is possible.  But the cut
between the previous fantastic sequence of successful predictions
\emph{right on the mark}, and, on the other hand, the later series of
unsuccesses is striking.  Before, experimental particle physicists
were always smiling and walking like heroes: looked like God was
reading Phys Rev D and implementing all suggestions of the theorists. 
Nowadays, thanks god they are still busy figuring out aspects of the
standard model, because all the new physics that theoretician have
suggested wasn't there \ldots

\si Theory has always made wrong predictions. 

\sa Yes, but also right predictions, and those are missing, after the
standard model.

\si It is because energies of new predicted physics are too high. 

\sa Not at all.  There have been plenty of predictions that were well
within reach.  They just were wrong. 

\si So, what do you make of this?

\sa That perhaps Nature is telling us that our path of theoretical
research has taken a wrong turn, at some point \ldots

\si This is not a proof. 

\sa Of course.  The fact is that we do not know.  But it is, to say
the least, a strong reason for exploring alternatives to what you
called ``the natural path of the evolution of high energy physics". 
It is a strong reason for being suspicious of the idea that string
theory has to be right just because this path got to it.  One follows
a path with confidence as long as indications are positive; why should
we all keep following a path, altogether, when negative indications
pile up?

\si Maybe \ldots\ But what if supersymmetry is seen?

\sa Then we'll have another conversation.  But I have heard so many
announcements that supersymmetry is ``on the verge of being seen".  I
am told that famous theoreticians claimed that supersymmetry was
certainly going to be found in a year or two, otherwise they'd change
their mind.  This was many years ago, and they haven't changed their
mind yet.  I understand changing mind is hard, especially just because
of experimental evidence\ldots \ But do we believe nature or we
believe our fancies?  I remember myself one ``very" famous theoretical
physicist giving a major talk in front of a big audience of
mathematicians and saying that his experimental friends had just told
him that the first evidence for supersymmetry was showing up in the
data\ldots he gave it as a great announcement\ldots\ everybody was
thrilled\ldots \ In the same talk, he also announced that what
mathematicians will do in the coming millenium is to study string
theory\ldots\ 

\si Sal, no sarcasm\ldots

\sa Alright, I apologize.  Let me come to another point of yours. 
That strings bring everything together, it is a theory of everything.

\si You cannot deny that. 

\sa No, I don't deny that.  But I am not sure that running
after the theory of everything is the right way to go.

\si Do you think that an attempt to merge separate theories is 
misled? 

\sa Not al all, this has been very effective historically.  What I am
disputing is the idea of the theory of everything.

\si It is the old dream of physics.

\sa Yes, but it has never worked.  And it might fail to work this time
as well.

\si This time is different.  We have theories that almost explain
everything we see.

\sa This time is not at all different than the other times. 
Physicists have repeatedly believed in the past that they had theories
that explain ``almost everything we see".  The feeling that we are
``almost" at the theory of everything was there just before quantum
theory, at the time of Maxwell, just after Newton \ldots it has always
been wrong \ldots

\si I am not a historian.  It may be right this time \ldots

\sa On the basis of which evidence? 

\si String theory \ldots 

\sa A theory, we agreed, that so far does \emph{not} describe the
world we live in, does \emph{not} give any precise univocal
prediction, and, I can add, whose general foundations are still
completely \emph{un}clear?

\si Wait a minute.  The theory is not in such a bad shape.  The
perturbative theory allows us to compute all finite scattering
amplitudes in the deep quantum gravitational regime.

\sa Does it really ?  Quantum gravitation regime is when 
center of mass energy is far above Planck.

\si And? 

\sa And this is where the perturbation expansion stops converging
\ldots

\si You mean the divergence of the perturbative series
itself, not the infinities in the individual terms.

\sa Yes. 

\si But the series diverges in all quantum field theories. 

\sa Yes, but those are known to be approximations.  You can rely on
some other theory at higher energy.  This is supposed to be the final
theory \ldots\ The fundamental theory does not allow us to compute at
the Planck scale?

\si Okay, not the perturbative theory \ldots but then there are the
nonperturbative aspects of the theory \ldots\ In some cases it is
possible to give a non-perturbative prescription for the definition of
amplitudes, such as matrix theory for 11d space or AdS/CFT for
asymptotically AdS spacetimes.

\sa Anything about \emph{our} world? 

\si No sarcasm, Sal.  You cannot discount all the nonperturbative 
aspects of the theory. 

\sa You mean the dualities, the various maps between strong coupling
and weak coupling, Joe Polchinski's branes and all that \ldots

\si yes, the theory is far more rich that we expected, 
it is fantastic how \ldots

\sa I know, I go to seminars and hear the exclamations \ldots 

\si So? 

\sa So what? 

\si So you are not convinced by that?

\sa What should I be convinced of? 

\si That we are beginning to understand the non perturbative regime as
well, and remarkable phenomena happen.

\sa Are you saying that the non perturbative regime is understood? 
That we can routinely compute in the nonperturbative regime?

\si Far from that.

\sa So, the theory does \emph{not} allow us to compute finite
scattering amplitudes in the deep quantum gravitational regime\ldots
Would you agree in saying that the theory is well understood
perturbatively, where it does \emph{not} look like the real world, and
we have only glimpses on its nonperturbative regime, but not yet a
clear relation with our world?

\si I guess I would.

\sa Well, after having been developed for so many years by the
smartest physicists on the planet, numbered by the hundreds \ldots\
this sound quite a poor result to truly excite me \ldots

\si Your taste \ldots\ Remains the solid fact that the theory provides
a finite perturbation expansion for quantum gravity.

\sa Fair.  And this is remarkable, I agree.  But even on that I have
doubts.

\si Doubts? 

\sa Is there a proof that the theory is finite at all orders?

\si Everybody says is finite. 

\sa Everybody says so.  But does anybody know for sure?

\si There are many indications. 

\sa Many indications is different from knowing for sure.  There were
also indications that supergravity was finite at all orders, and
famous physicists gave inspired talks that the final theory of
everything was found.  It turned out not to be finite, at three loops
or something.

\si hmm\ldots

\sa Let me put it short: is there a paper, a book, a report, which
shows that it is finite at all orders?  I am not asking for something
convincing a mathematician.  Just something convincing a field
theorist who is just a bit skeptical.  In 1986 the book by Green,
Schwarz and Witten said that finiteness to all order is a common belief
among string theorists, but complete proofs had not appeared yet.  Now
is more than fifteen years later.  Have they since?

\si There is a 1992 paper by Mandelstam \ldots 

\sa I know that paper.  It proves that the divergence that people were
most worried about, the dilaton divergence, does not occur in
superstrings; but there are other sources of divergence.  The books by
Kaky and Polchinski, written after that paper, write quite clearly that
there is no proof\ldots

\si I do not really know if a proof of finiteness at all orders exists
\ldots\

\sa I have tried to found out.  Perturbative finiteness was never
shown past two loops.  Actually, it is not even known if there is an
unambiguous prescription for writing superstring amplitudes past genus
2.  It is not clear if there is any well defined theory there.  I've
not spoken to anyone who is optimistic about the probability of the
general case, except people who tell me its been proved long ago in
some obscure paper only they know about, but can't recall the exact
reference, and never send it after promising to.

\sa Listen, you have studied a bit of strings: it is a beautiful and
vast world.

\sa Yes, but ``\ldots our arguments have to be about the world we
experience, not about a world made of paper".

\si It is not just a world made of paper.  The theory predicts
fermions, gauge fields, quantum theory, and especially it predicts
gravity.  In a world where gravity was not observed, a theoretician 
with string theory would have predicted the existence of gravity. 

\sa Prof, do you really believe this? 

\si Well, maybe no.

\sa In a world where gravity was not observed, a theoretician, having
noticed that the Veneziano amplitude disagrees with reality would have
just discarded it.  The reason we all got interested in string theory
is because there is gravity in it, without previous knowledge of
gravity, string theory would not have been taken seriously.  I can
write a theory of the standard model plus a field called Pippo, where
the Pippo field could not exist without the Standard model, and then
claim: ``look!  my theory is great: if we didn't know the standard
model, my theory would have discovered it!  therefore my theory is
right!  therefore the Pippo field exists".  It is obviously a
nonsense.  We develop only the theories that agree with what we know
so far.  It is silly then to be proud that they agree with what we
have seen so far.  It is as if Weinberg claimed that the $SU(2)\times
U(1)$ theory ``predicted electromagnetism", and in a world where
electromagnetism had not been observed, he would have predicted it. 
It is a nonsense.  In a world without electromagnetism he would not
have invented his theory.  In fact, Weinberg and Salam and Glashow
never claimed that about their theory.  The remarkable prediction of
their theory, which gave confidence in it, was the neutral currents
and the W and Z particles\ldots\ Making a big deal of the fact that
there is gravity in string theory is the kind of argument that can be
raised only out of the desperation that this theory is not able to
predict anything new with certainty\ldots

\si I guess many people would agree with this\ldots 

\sa This leaves the last point: that string theory is the only known
way of combining GR and QM. Which leads us to loops.

\si Enough strings? 

\sa Yes, your turn to attack \ldots Attack is easier than
defense, given that we do not have any experimentally proven theory
yet \ldots

\si Alright.  I know little on loop gravity, so, correct me if I am
wrong.  But for what I hear, the theory has difficulties in recovering
the low energy limit.

\sa True.  It might be doable, but it is not done.  One can write
states related to certain classical solutions, but there is not yet a
way to recover low energy perturbation theory.

\si And there isn't a single loop quantum gravity. 

\sa You mean the definition of the hamiltonian constraint
admits many variants.  True.  

\si Alright, this is the incompleteness, and I suppose smart people
like you think they may fix it \ldots

\sa Thanks for the ``smart", Prof.  But you said no sarcasm\ldots!

\si Alright!  Let me come to the serious points.  First, we know there
is no way of combining GR and QM without altering GR or adding matter.

\sa How do we know this? 

\si Because GR is nonrenormalizable. 

\sa This does not mean anything.  There are several examples
of quantum field theories that are well defined nonperturbatively, and
are nevertheless nonrenormalizable if we try a perturbation expansion.

\si But why should GR be like that?  GR is like Fermi theory. 
Empirically successful but nonrenormalizable.  Therefore we must
change its high energy behavior, like we did with Fermi theory.

\sa How can you be sure that GR is like Fermi theory?  This is one
possibility, of course, but there is another possibility: that GR is
\emph{not} like Fermi theory, and that the reason it is
nonrenormalizable is a different one.

\si Which one? 

\sa That it is the weak field perturbation expansion that 
fails for GR. 

\si Why should it? 

\sa Because the weak field perturbation expansion is based on Feynman
integrals that sum over infinite momenta, namely over regions of
arbitrary small volume.

\si So? 

\sa Simple dimensional arguments show that these regions are
unphysical in quantum gravity.  They literally do not exist.  It make
no sense to integrate over degrees of freedom far smaller than the
Planck length.  In fact, loop gravity strongly supports this
possibility, because one of the results of the theory is that volume
is discrete at the Planck length.  There is literally no volume
smaller than a Planck volume in the theory.

\si I suppose this is an hypothesis of the theory.

\sa No, it is not an hypothesis, it is a result. 

\si How can that be? 

\sa The volume is a function of the metric, namely of the
gravitational field.

\si Okay.

\sa And the gravitational field is quantized. 

\si Okay.

\sa Therefore the volume is a quantum variable. 

\si I am following. 

\sa Therefore it may be quantized. 

\si And how do we know if it is? 

\sa As usual in quantum theory: we compute the spectrum of the
corresponding operator.

\si You mean like the energy of an harmonic oscillator? 

\sa Precisely. 

\si And? 

\sa And the calculation shows that the spectrum is discrete and there
is a minimal nonvanishing volume.  Hence in the theory there can be no
Feynman integral over arbitrarily small volumes.

\si I am a bit confused.  If the discrete volume is a result and not
an input, what is the physical spacetime on which the theory is
defined?

\sa There is none.

\si I do not understand. 

\sa It is a background independent formulation. 

\si But how can a field theory ``not be defined on a spacetime".

\sa This is precisely what happens in classical GR.  In 
fact, this is background independence realized. 

\si In GR things move on a spacetime.  Fields and particles
have a dynamics on a curved spacetime.  Maybe curved, but always a
spacetime.

\sa Physics on a curved spacetime is not GR. GR is the dynamics of
spacetime itself.  So, quantum GR is the theory of a quantum
spacetime, not a quantum theory on various spacetimes.

\si But how can we do physics without a spacetime?  You will
not have energy, momenta, positions \ldots

\sa Indeed. 

\si We do not know how to do physics without these concepts.

\sa General relativistic physics, theoretical and observational, does
very well without.  Energy, momenta and positions can only be defined
in certain limits or relative to certain objects.

\si But this means changing all the basic tools of quantum field
theory.

\sa This is precisely what happens in loop quantum gravity. 

\si Wait, all our experience in quantum field theory teaches us that
these tools are essential.  Quantum field is the most effective tool
we have to understand the world.  I am not ready to abandon it.

\sa But GR teaches us that we should do so. 

\si You take GR too seriously.  GR is just an effective nonlinear
lagrangian for describing the gravitational interaction.  It is most
presumably just a low energy lagrangian.  I would be surprised if
there are no high energy corrections to the Einstein Hilbert action.

\sa I think there is a confusion here. 

\si A confusion? 

\sa Yes, between the details of the Einstein-Hilbert action and the
general lesson of GR, which is diffeomorphism invariance or background
independence.  When the loop people talk about taking GR seriously, or
about the lesson of GR, they do not mean the specific form of the
Einstein-Hilbert action.  They mean the fact that the fundamental
physical theory has to be background independent.  This means that in
the fundamental theory there isn't a fixed background spacetime and
fields on it.  There are just many fields that construct the spacetime
itself.  This is the conceptual novelty of GR that the loop people
want to merge with quantum field theory.  Not the details of the
Einstein Hilbert action.

\si But background independence is also what string theory people are
trying to achieve.

\sa Yes, the question is why trying to achieve this with all the
gigantic apparatus of string theory --and nobody seems to be
succeeding yet-- when it seems to be possible with just conventional
GR alone --and the loop people seem already to be succeeding\ldots

\si There are many indications in string theory that the 
background independent theory exists. The various dualities connect 
different expansions, they are all aspects of the same theory \ldots 

\sa But nobody knows the general background independent formulation of
this hypothetical theory \ldots

\si Yes indeed. 

\sa While in loop gravity the background independent formulation is
known.

\si But in this funny way without spacetime, without energy, without
momenta, without all the usual machinery.

\sa Everybody says they want background independence, and then when
they see it they are scared to death by how strange it is \ldots\
Background independence is a big conceptual jump.  You cannot get it
for cheap, with conventional means.

\si You can define string theory nonperturbatively by means of a flat
space theory defined on the boundary of spacetime.

\sa Yes, Juan Maldacena has indicated the way.  But his model does not
describe our world, it is highly unrealistic\ldots 

\si \ldots yes, but it indicates that there may be a possibility of
defining a realistic background independent theory via a boundary 
theory. 

\sa Maybe, but I haven't seen the realistic model yet.  It might be
certain bulk theories are related to certain boundary theories,
perhaps because they have the same symmetries, or something; or maybe
are related in some sectors, I do not know.  But even if it was true
that a certain background independent theory can be mapped on a flat
space theory, can we say we have understood background independent
physics?  You can map a special relativistic theory over a theory with
a preferred frame, and compute there.  As far as you do that, you have
not yet understood Lorentz invariant physics\ldots\ We want to find
the right way of thinking in the background independent regime, not
just map ourselves out of it.

\si Of course, but this might be a useful first step.

\sa Of course, very well.  Far from me to deny that in string there is
an active search for background independent physics, or hints and
glimpses.  What I am saying is that in loop gravity there is already
background independence, fully realized in the basics of the theory.

\si I may concede this, but in exchange one cannot go down to low
energy physics.  If loop theory is correct, can you compute the cross
section for graviton-graviton scattering?  Using your finite minimal
volume, can you fix all the constants in front of the terms that
conventional perturbation theory leaves undetermined?

\sa I think they are working on it, but I haven't seen anything solid
yet\ldots\ I guess this is the weak point of the loops, right now
\ldots

\si Good.  I agreed about so many weaknesses of string theory!

\sa Fine \ldots

\si So, suppose I believe your quantum theory of general relativity 
alone without matter.  Then I am still very far from a realistic 
theory. There is matter in the world. 

\sa In loop gravity you can easily couple fermions and Yang Mills
fields.  In fact, you can even do a supersymmetric theory if you want,
the point is that it is not required by consistency, it is not
required by experience, so there is not much interest.  There were a
few papers indicating it is possible.  So, you just couple the matter
you see in the world to quantum GR.

\si And you have no explanation of why there is that particular form 
of matter, that particular coupling in the standard model. 

\sa No.  But so far strings do not seem to be more successful at that
either.  Hoping that some nonperturbative physics that we do not yet
understand will pick the right Calabi-Yau manifold out of a million is
not so much better than honestly saying that we do not understand why
$SU(3)\times SU(2)\times U(1)$.  I think we are still very far from
the end of physics!  Which is good, for us young people\ldots\ Who
knows, we simply do not understand the deeper physical reason of the
standard model.  I find a more appealing explanation of the standard
model in Alain Connes vision, which ties it to a simple underlying
geometry, than in the string idea that it is the minimum of a
potential we know nothing about.

\si I suppose when you add matter to loop gravity you loose finiteness
because there you get the usual infinities back.

\sa Not at all!  In fact finiteness extends even to, say QCD coupled
to gravity.  For the very same reason: there is no small volume.  You
see, from the point of view of QCD, being coupled to gravity is very
much like living on a Planck scale lattice, where the theory has no
infinities.  

\si So, what is precisely the status of these finiteness results? 

\sa There are two sorts of finiteness results, as far as I have
understood.  In the hamiltonian formulation of the theory, one proves
that the operators that define the theory nonperturbatively do not
develop divergences.  In fact, the mathematical foundations of loop
gravity are extremely solid.  They have been developed to the level of
rigor of mathematical physics.

\si I know.  On the one hand, this has made the theory solid, but on
the other hand, it has made the language harder to follow for a high
energy physicist.

\sa Then, there is also another formulation of the theory, called
spinfoams, which is a Feynman-like perturbation expansion for
computing amplitudes.  At least for some euclidean versions of this,
there are mathematical theorems that state that the expansion is
finite.

\si Up to which order in perturbation expansion? 

\sa At any order. 

\si You mean there is a perturbation expansion which has been truly
proven to be finite al all orders?

\sa Yes sir. Unlikely for strings. 

\si So, why one cannot compute all scattering amplitudes, say between
gravitons, out of this?

\sa Because the expansion is defined in a certain basis, and one does
not know yet how to write the Minkowski vacuum state and the graviton
states in this basis\ldots 

\si I see \ldots\ for a moment you almost convinced me to study loops
\ldots\ So, if the loop people are not yet able to describe gravitons,
what sort of physics can they actually describe, besides providing
this loopy picture of reality at the Planck scale?

\sa Black holes, with their entropy, early cosmology\ldots 

\si Yes, I have heard there is an active ``loop quantum cosmology",
even with claims that inflation could be driven by quantum gravity
effects\ldots \ But let me come to some serious objections.  Is it true
that the Hilbert space of the theory is nonseparable?

\sa No, it is not true.  At some stage it was not properly defined. 
There is now a proper definition where the Hilbert space is separable.

\si But the theory is based on loop states, which are created by
holonomy operators \ldots

\sa yes

\si \ldots and we know that in QCD these states are not good.  They
are nonnormalizable; the field operator is smeared only in one
dimension, which is not enough.  And if you try to take these states
as orthogonal basis states, you get everything wrong.  It is the 
very starting point of the loop representation that is wrong. 

\sa All you said is correct in QCD.   But gravity is different. It is 
truly different. 

\si Why? 

\sa Precisely because of diff-invariance.  Or, if you want, because
the volume is quantized.  Physically, it is not that the loop states
are concentrated on infinitely thin lines: it is as if they have a
Planck size.  What happens mathematically is that the localization of
the loop on coordinate space is pure gauge.  The physical degrees of
freedom are not in the localization of the loop, but only in what
remains after you factor away diffeomorphisms, namely they are in the
way the loops intersect or link.  In fact, the infinities are
precisely washed away by diff-invariance.

\si I am not sure I understand this. 

\sa Well, you have to enter the math of the theory. But the point is 
that the loop states become good states in gravity.  Let me put it in 
this way.  On a lattice, loop states form a perfectly well defined 
basis, right?

\si Yes of course, it is the continuum limit that gives problems. 

\sa Well, in gravity is like being always on a Planck size lattice,
because each loop state does not live on a background spacetime.  It
lives on the lattice formed by all the other loops.

\si Hmm.  I vaguely see.  Is the theory Lorentz invariant?

\sa I have no idea.  I suppose that it is just like in classical GR.
Lorentz invariance is not broken if the state of the gravitational
field happens to be Lorentz invariant, and is broken if it is
not\ldots

\si You are confusing the symmetry of a solution with the
symmetry of the theory.  Classical GR is Lorentz invariant. 

\sa No, it is not.  The Lorentz group acts in the tangent space at
each spacetime point, of course.  But the theory is not Lorentz
invariant in the sense you mean.  If it were, we could Lorentz
transform all solutions of the theory, right?  \ldots like we can
Lorentz transform all solutions of Maxwell theory.

\si We can't? 

\sa What do you get if you boost of a cosmological Friedmann
solution?

\si Okay, you are right.  But if we additionally assume that 
spacetime is asymptotically Minkowskian\ldots

\sa Then yes, of course, if you impose Lorentz invariant
boundary conditions you introduce a Lorentz invariance into the
theory.  There will be the asymptotic Lorentz group acting\ldots But I
am not sure there are quantum states that are exactly asymptotic
Minkowskian in the quantum theory.  Perhaps there are, perhaps at the
Planck scale the symmetry is spontaneously broken by the short scale
structure.  Like a given crystal breaks the rotational symmetry of the
dynamical theory of its atoms.  But I don't really know\ldots

\si But isn't the existence of a minimal length obviously
intrinsically incompatible with Lorentz invariance?

\sa No, this is a misconception. 

\si Why?  If I slowly boost the minimal length, it smoothly
becomes shorter\ldots

\sa No, this is quantum theory.  It would be like saying that
the existence of a minimal size of the z-component of the angular
momentum breaks rotation invariance, because you can smoothly rotate
it to zero.  In quantum theory what changes smoothly is the
probability of getting this or that eigenvalue, not the eigenvalues
themselves.  Same with minimal length, which appears as an eigenvalue. 
If you begin boosting something which is in a length eigenstate, you
get a smoothly increasing nonvanishing probability of getting a
different length eigenvalue, not a shorter eigenvalue.

\si Ah!  Nice.  So, does loop gravity predicts Lorentz violation or 
not?  

\sa I am not sure.  I think so far it is like large extra dimensions
for strings.  Could be.  Could not.

\si hmm\ldots\ But if you have no Lorentz symmetry, you may have no
hermitian hamiltonian.  Is loop gravity unitary?

\sa No, as far as I understand. 

\si This is devastating. 

\sa Why?  

\si Because unitarity is needed for consistency.

\sa Why? 

\si Because without unitarity probability is not conserved. 

\sa Conserved in what? 

\si In time.

\sa Which time? 

\si What do you mean ``which time?".  Time.

\sa There isn't a unique notion of time in GR. 

\si There is no coordinate t? 

\sa There is, but any observable is invariant under change of t,
therefore everything is constant in this t just by gauge invariance.

\si I am confused. 

\sa I know, it is always confusing\ldots Nonperturbative GR
is quite different from physics on Minkowski \ldots

\si Do we really need to get in the conceptual complications of 
GR? 

\sa Well, if we are discussing the theory that is supposed to 
merge GR and QM \ldots 

\si String theory merges the two without these complications. 

\sa This is why I think that string theory does not really 
merge GR and QM. 

\si But you agreed it does.

\sa No, I agreed that strings provide a finite perturbation
expansion for the quantum gravitational field and this expansion
breaks down when things begin to become interesting: in the strong
field regime.

\si So, why does string theory not merge GR and QM? 

\sa Precisely because GR tells us that there is no fixed
background space with stuff over it.  Strings are always about
background spaces with stuff over them.

\si But the background derives simply from the split between
the unperturbed and the perturbed configuration of the field, which we
always do in quantum field theory.   

\sa We do so in perturbative theory.  We do not do so if we define QCD
as the limit of the lattice theory.  And weak field perturbation theory
might not work in gravity.

\si And what do you make of what is known about nonperturbative string
theory?

\sa What is known are relations between theories defined over
different backgrounds.  These are hints that the background
independent theory might exist.  But this is far from understanding
the foundations of the background independent theory.

\si The fully background independent theory is an immense task, we are
far from it.

\sa Loop gravity does it.

\si And what do field and things stand on?

\sa On top of each other, so to say. 

\si It is not quite similar to the physics I know. 

\sa It is beautiful.  You talked about the beauty of string theory. 
The emergence of spacetime as excited states, as loop and spinnetwork
states, is extremely beautiful.  It is quantum theory and general
relativity truly talking to each other \ldots

\si If background spacetime is missing, so is time? 

\sa Yes sir. 

\si And if you do not want to impose asymptotic flatness you do not
have asymptotic background time either?

\sa Yes sir. 

\si And if there is no background time, there cannot be unitary
evolution, right?

\sa Yes. 

\si I am not sure I can digest a theory where there is no space and no
time to start with, and without unitarity\ldots

\sa I suppose this is why there is so much resistance to loop gravity
\ldots\ Again, everybody searches background independence, but when
you see it, it is sort of scary\ldots\ Anyway, we can all believe
what we like, until experiments will prove somebody right and somebody
wrong, and for the moment no experiment is talking to us \ldots\
Future will tell \ldots\ But my point is that the absence of unitarity
does \emph{not} imply that the theory is inconsistent.  Only that the
notion of time is intertwined with dynamics.  It is similar to the
fact that there is no conserved energy in a closed universe \ldots

\si Alright, I accept this.  But we have been digressing\ldots\ can we
try to wrap up?

\sa Alright.  I suppose your conclusion is that loop gravity is (a)
too different from usual QFT, (b) not completed and (c) not yet able
to recover low energy physics\ldots

\si And your conclusion is that string theory (a) does not describe
the real world in which we live, (b) is not predictive because it can
agree with any experimental outcome, (c) it requires an immense
baggage of new phenomenology like supersymmetry, and extradimensions,
which we do not see, and (d) it has not lead to a true conceptual
merge of QM with the GR's notions of space time\ldots\ 

\sa Of course, they could both be wrong \ldots

\si Or both right: loops might end up describing some aspects of
quantum gravity and strings some other aspects\ldots

\sa Prof, maybe I was a bit carried away by the polemical verve, so
let me be clear.  I think that string theory is a wonderful theory.  I
have a tremendous admiration for the people that have been able to
build it.  Still, a theory can be awesome, and physically wrong.  The
history of science is full of beautiful ideas that turned out to be
wrong.  The awe for the math should not blind us.  In spite of the
tremendous mental power of the people working in it, in spite of the
string revolutions and the excitement and the hype, years go by and
the theory isn't delivering physics.  All the key problems remain wide
open.  The connection with reality becomes more and more remote.  All
physical predictions derived from the theory have been contradicted by
the experiments.  I don't think that the old claim that string theory
is such a successful quantum theory of gravity holds anymore.  Today,
if too many theoreticians do strings, there is the very concrete risk
that all this tremendous mental power, the intelligence of a
generation, is wasted following a beautiful but empty fantasy.  There
are alternatives, and these must be taken seriously.  Loop gravity is
pursued by a far smaller crowd; has problems as well, as you pointed
out, but is succeeding in places where strings couldn't get, and is
closer to reality.  And if you think at the quantum excitations
bulding up physical space, you truly see quantum mechanics and general
relativity talking to one another.  And is beautiful.  I have an
immense respect for string theorists, but I think it is time to
explore something else.  Don't you think, to say the least, that both
theories are worthwhile exploring?

\si \ldots

\vskip1cm

{\em The final words of Professor Simp were not heard.  But he was
seen smiling, and later heard referring to Sal as stubborn, but
definitely smart.  By the way, Sal is still looking for a 
job\ldots}

\vskip2cm
\rightline{\today}

\vfil
\centerline{-----------}
\noindent{\small \em Many thanks to Ted Newman, Gary Horowitz, Abhay
Ashtekar, Daniele Oriti, Lee Smolin, Warren Siegel, Simone Speziale
and Juan Maldacena for corrections and suggestions.}

\end{document}